\documentclass[times,ShortAfour,sagev]{sagej}

\usepackage{moreverb,url}

\begin{document}

\runninghead{F. Palm\'i-Perales et al.}

\title{Approximate Bayesian inference for multivariate point pattern analysis in disease mapping}

\author{Francisco Palm\'i-Perales\affilnum{1}$^*$, Virgilio G\'omez-Rubio\affilnum{1}$^*$, Gonzalo L\'opez-Abente\affilnum{2,3}, Rebeca Ramis-Prieto\affilnum{2,3}, Jos\'e Miguel Sanz-Anquela\affilnum{4} and Pablo Fern\'andez-Navarro\affilnum{2,3}}

\affiliation{\affilnum{1}Department of Mathematics, Universidad de Castilla-La Mancha, Albacete, Spain\\
\affilnum{2}Environmental and Cancer Epidemiology Unit, Carlos III Institute of Health, Madrid, Spain\\
\affilnum{3}Consortium for Biomedical Research in Epidemiology \& Public Health, CIBER Epidemiolog\'ia y Salud P\'ublica - CIBERESP, Spain\\
\affilnum{4}Cancer Registry and Pathology Department, Hospital Universitario Pr\'incipe de Asturias and Department of Medicine and Medical Specialties, Faculty of Medicine, University of Alcal\'a, Alcal\'a de Henares, Madrid, Spain}

\corrauth{Francisco Palm\'i-Perales and Virgilio G\'omez-Rubio, Department of Mathematics, School of
Industrial Engineering, University of Castilla-La Mancha, 02071, Albacete,
Spain}
\email{\{Francisco.Palmi,Virgilio.Gomez\}@uclm.es}

\begin{abstract}

We present a novel approach for the analysis of multivariate case-control
georeferenced data using Bayesian inference in the context of disease mapping,
where the spatial distribution of different types of cancers is analyzed.
Extending other methodology in point pattern analysis, we propose a log-Gaussian
Cox process for point pattern of cases and the controls, which accounts for
risk factors, such as exposure to pollution sources, and includes a term to
measure spatial residual variation.

For each disease, its intensity is modeled on a baseline spatial effect
(estimated from both controls and cases), a disease-specific spatial term and
the effects on covariates that account for risk factors.  By fitting these
models the effect of the covariates on the set of cases can be assessed, and
the residual spatial terms can be easily compared to detect areas of high risk
not explained by the covariates. 

Three different types of effects to model exposure to pollution sources are
considered. First of all, a fixed effect on the distance to the source. Next, smooth terms on the distance 
are used to model non-linear effects by means of a discrete random walk of
order one and a Gaussian process in one dimension with a Mat\'ern covariance.
Spatial terms are modeled using a Gaussian process in two dimensions with a
Mat\'ern covariance.

Models are fit using the integrated nested Laplace approximation (INLA) so
that the spatial terms are approximated using an approach based on solving
Stochastic Partial Differential Equations (SPDE).  Finally, this new framework
is applied to a dataset of three different types of cancer and a set of
controls from Alcal\'a de Henares (Madrid, Spain). Covariates available
include the distance to several polluting industries and socioeconomic
indicators. Our findings point to a possible risk increase due to the proximity to some
of these industries.

\end{abstract}


\maketitle

\section{Introduction}

The analysis of point patterns plays an important role in Public Health.
Case-control studies are often conducted to assess whether the spatial
distribution of the cases follows that of the control or a different pattern
caused by risk factors, such as exposure to pollution sources.  The use of
a set of controls in the analysis of the locations of cases of a disease is
important for two reasons. First of all, it allows for adjusting for the
spatial distribution of the population. Secondly, by comparing the cases and
the controls it is possible to identify risk factors associated to the disease.

The first topic is often referred as the study of the spatial risk
variation. For point patterns, it is often common to take the
ratio of the intensities of cases and controls\cite{Diggle:2003}. The study
of this ratio can also be of interest in order to detect local
hotspots or areas where the intensity of the cases is large, even
after accounting for the spatial distribution of the population and
other risk factors\cite{diggle2007differences}.

Assessing risk factors is often based on covariates associated to the cases and
the controls. Although it is common to find socio-economic covariates, it is also
possible to find covariates that measure exposure to putative pollution
sources\cite{DiggleRowlingson:1994}.  A common proxy for exposure is the
distance to the pollution sources, which is often easy to compute for point
patterns analysis when the locations of the pollution sources are known.

Modeling the intensity of the cases and the controls can be approached in a
number of ways\cite{gelfand2010handbook,baddeley2015spatial}. First of all, if
both patterns are considered separately, a non-parametric estimate can be
obtained by means of kernel smoothing and similar methods.  If covariates are
available, these can be included in the estimation of the intensity by means of
a log-Gaussian Cox process\cite{digglemoraga2013}.

In a case-control analysis, the intensity of the cases can be modeled
semi-parametrically by assuming that it is the intensity of the controls
modulated by the covariates\cite{diggle2007differences}. These estimates can
also be used to estimate the probability of being a case. When several types of
points are available, these methods can be extended so that a separate
intensity is estimated for each point type, and the distribution of the
probability of being a point of each type can be computed\cite{Diggleetal:2005}.

The study of spatial risk variation allows us to assess whether risk is
constant across the study region. This has often been conducted using Monte
Carlo random relabeling tests which are computationally
expensive\cite{Diggle:2003}.

In this paper we propose a new approach to the analysis of different diseases
using case-control data. For this, recent developments in Bayesian inference
and computational statistics will be used in order to extend the current
methodology to these new setting where the locations of cases of several diseases and a
set of controls are available. Models proposed will fall in the category of
log-Gaussian Cox processes where the log-intensity is modeled using the effect
of the covariates plus a shared spatial smooth term and disease-specific
spatial terms.

In particular, our models will be proposed within the framework of the
integrated nested Laplace approximation (INLA)\cite{INLA}.  INLA provides a
very flexible framework for model definition using different types of fixed and
random effects, state-of-the-art spatial models and computational speed.
Furthermore, the spatial smooth term will be modeled as a Gaussian process with
a Mat\'ern covariance, which will be approximated using the solution to a
Stochastic Partial Differential Equation (SPDE)\cite{SPDE,SPDELog-GausianCox}.
Socioeconomic factors will be included as fixed effects. Similarly, the effects
on the covariates that measure exposure to a pollution source will be considered
using a fixed effect, a smooth term using a random walk of order one and a
Gaussian process in one dimension with a Mat\'ern covariance (which will also
be approximated using a SPDE approach).

This new methodology will be applied to a real data set from Alcal\'a de
Henares (Madrid, Spain) on the locations of the cases of three types of cancer
(lung, stomach and kidney) and a set of controls. Furthermore, the locations of
different types of polluting industries are available.  Hence, the study
will be a case-control study to assess the spatial variation of the cases and
the relationship of the cases and the locations of the polluting industries
after accounting for the spatial distribution of the controls.

This paper is organized as follows. First, the real dataset from Alcal\'a de
Henares (Madrid, Spain) that motivated this work is introduced. Next, we
provide a summary of current approaches to the analysis of multivariate point
patterns to study disease risk variation and assessing exposure to pollution
sources, where our new methodological proposal is presented.  This is followed
by a description on how to use INLA for Bayesian inference on multivariate
point pattern analysis.  Finally, the methods described in this paper are
applied to the real data from Alcal\'a de Henares (Madrid, Spain). The paper
concludes with a discussion on the methods and results described herein.

\section{Cancer in Alcal\'a de Henares (Madrid, Spain)}

This work has been motivated by data obtained from Prince of Asturias
University Hospital (HUPA, Alcal\'a de Henares, Madrid, Spain). Cases have been
obtained from HUPA's Minimum Basic Data Set (MBDS)\cite{FernandezNavarroetal:2016,FernandezNavarroetal:2018}. The dataset obtained from
the MBDS contains cases of cancer of the lung (313), stomach (136) and kidney
(115).  Cases included people aged 40 or older diagnosed from January 2012 to
June 2014.  The set of controls is made of 3000 patients with non-cancer
diseases obtained from the MBDS, also from January 2012 to June 2014.  All
these three types of cancer have an important mortality
nationwide\cite{cancerspain:2009} and hence the interest of this study.

In addition to the case-control data, the locations of a number of polluting
industries in the Alcal\'a de Henares area have been obtained.  We used data on
industries governed by the IPPC and facilities pertaining to industrial
activities not subject to the IPPC Act 16/ 2002 but included in the E-PRTR
(IPPC + E-PRTR), provided by the Spanish Ministry for the Environment and Rural
\& Marine Habitats (Ministerio de Medio Ambiente y Medio Rural y Marino,
Spain). We selected the installations that where inside or close to the study
region. The geographic coordinates of their position recorded in the IPPC +
E-PRTR database were validated, by meticulously reviewing industrial
locations\citep{FernandezNavarroetal:2017}.

These have been provided by Health Institute 'Carlos III', that keeps a record
of all polluting industries in the country.  In particular, the locations of 13
air polluting industries are available, two of which can also be classified as
a heavy metals industries.  Other socio-economic variables at the census track
level are also available and provided by the Spanish Office for National
Statistics (INE).  

Because of the different nature of the types of cancers studied and that of the
polluting industries, it is likely that, in case there is any link between the
cases and the proximity to certain industries, different types of cancer will be
affected by different types of industries.

\begin{figure}
\centering
\includegraphics[width=6cm]{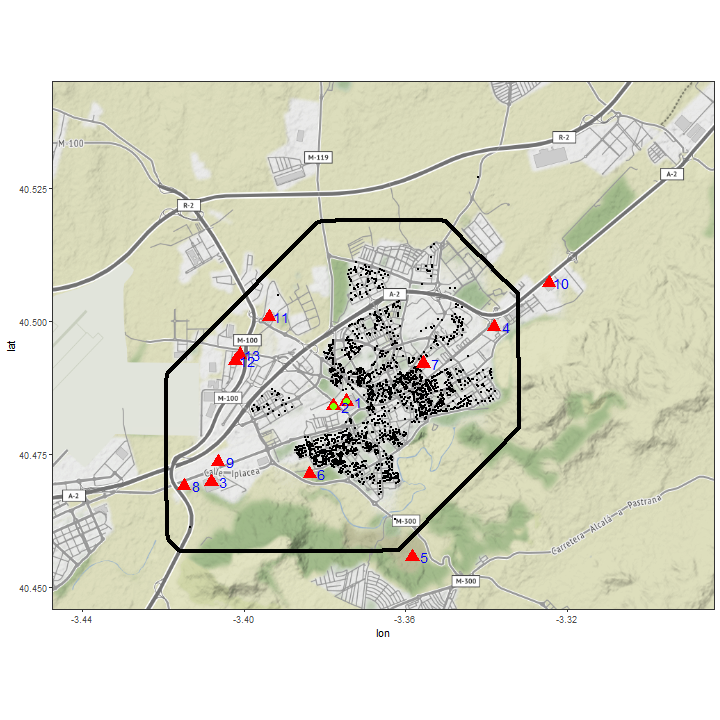}
\includegraphics[width=6cm]{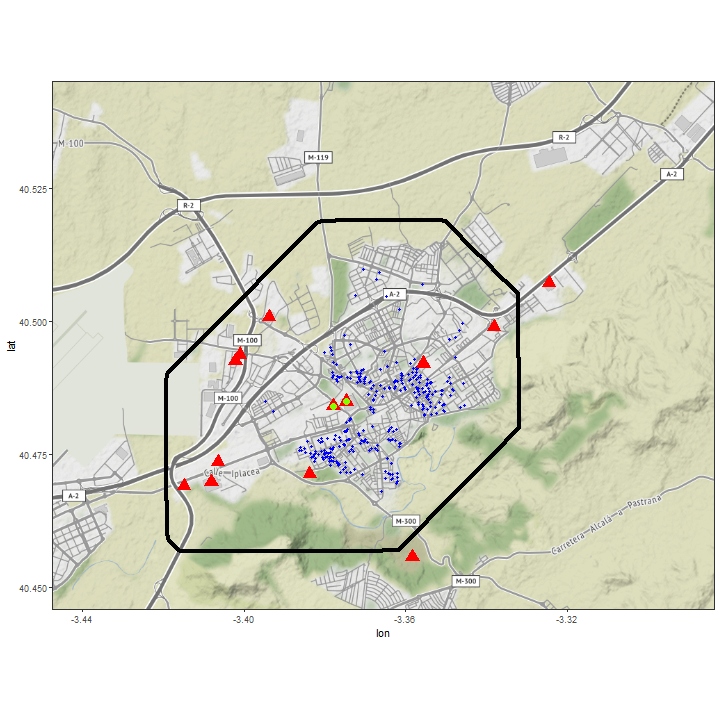}
\includegraphics[width=6cm]{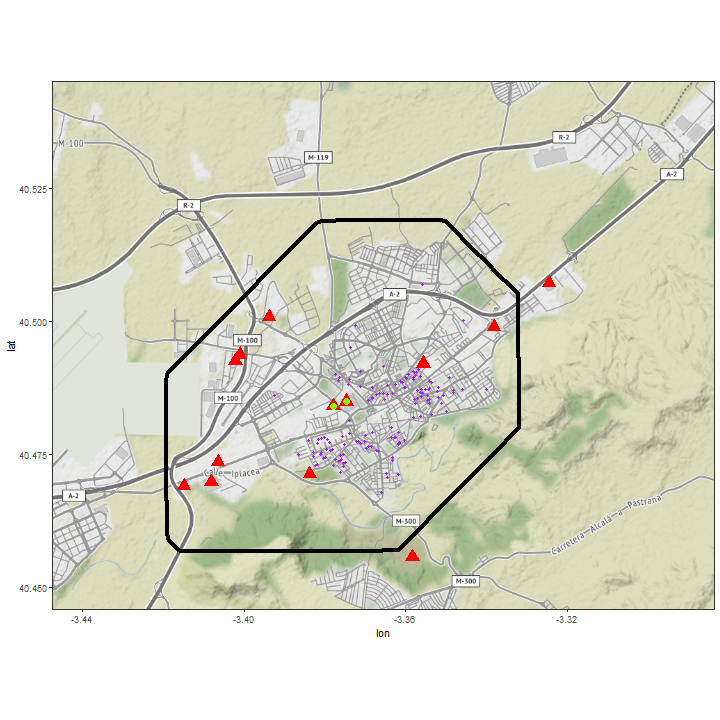}
\includegraphics[width=6cm]{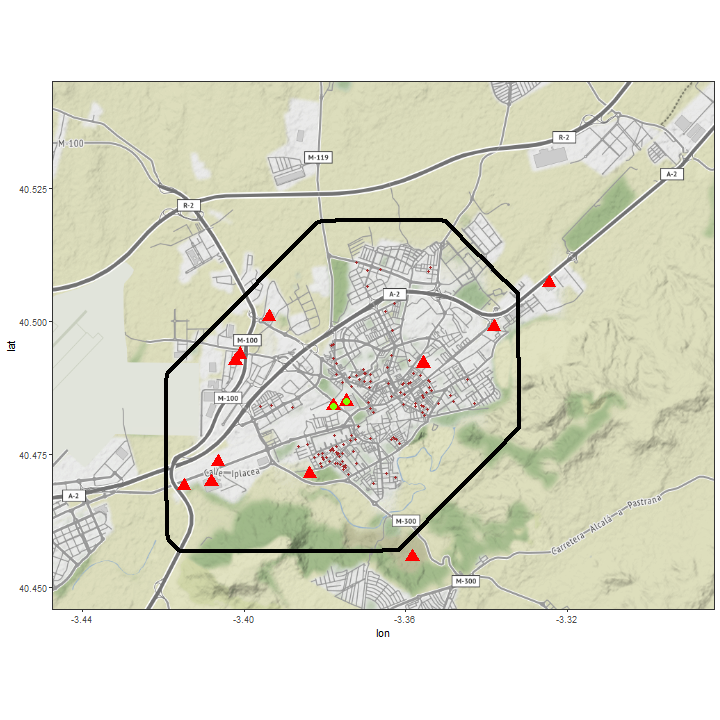}
\caption{From top to bottom and left to right, locations of the controls and
cases of lung, stomach and kidney cancer, respectively. The red triangles
represent the locations of air polluting industries and green dots the
locations of the heavy metals industries. The boundary represents the study region.}
\label{fig:locations}
\end{figure}

Figure~\ref{fig:locations} shows the locations of the cases and controls, as
well as the locations of the polluting industries. It is clear that, while
most of the polluting industries are outside the city, a few of them remain
close to the city center.

An inhomogeneous Poisson point process will be assumed for each point type.
The intensity of the controls at a location $x$ of the study area $\mathcal{D}$
will be represented by $\lambda_0(x)$, while the intensities of the cases of
lung, stomach and kidney cancer will be represented by $\lambda_i(x), i=1,2,3$,
respectively.  Similarly, $n_i,i=0,\ldots,3$ will represent the number of
controls and cases of the different types of cancer.

A simple way to assess spatial variation is to compute ratios $\rho_i(x) =
\lambda_i(x) / \lambda_0(x)$\cite{KelsallDiggle:1995SiM}.  In the case of no
risk variation, the distribution of cases will follow that of the population,
i.e., $\lambda_i(x) = \frac{n_i}{n_0}\lambda_0(x),i=1,2,3$.  For this reason,
under no spatial risk variation, ratio $\rho_i(x)$ will be equal to $n_i/n_0$.

In practice, the intensities involved need to be estimated, and these will be
denoted by $\hat\lambda_i(x),i=0,\ldots,3$. A simple and popular estimate can
be obtained with kernel smoothing\cite{Diggle:1985}. Note that in order to
estimate the ratio of the intensities the same bandwidth of the kernel must be
used\cite{KelsallDiggle:1995Bern}.

\begin{figure}
\centering
\includegraphics[width=2.95cm]{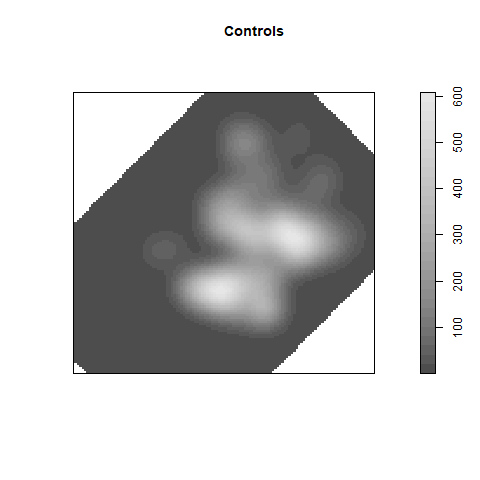}
\includegraphics[width=2.95cm]{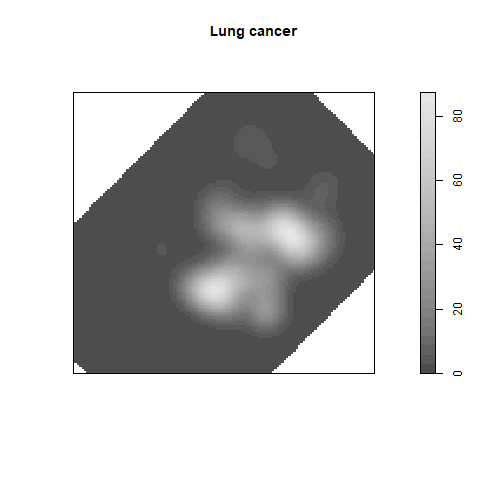}
\includegraphics[width=2.95cm]{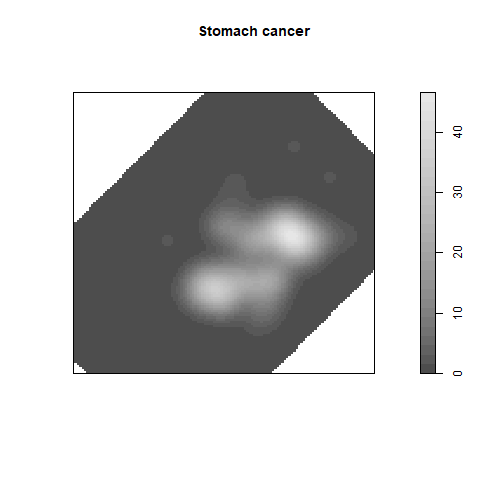}
\includegraphics[width=2.95cm]{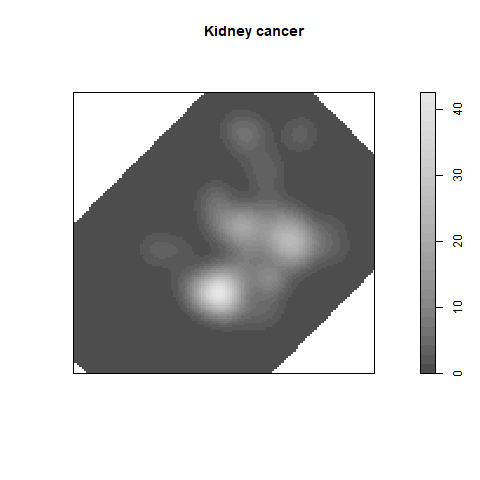}
\includegraphics[width=3.75cm]{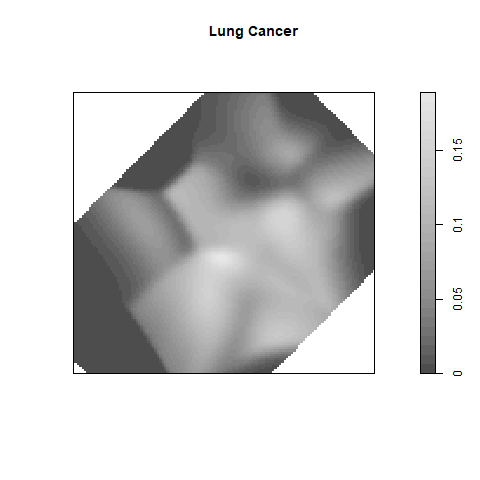}
\includegraphics[width=3.75cm]{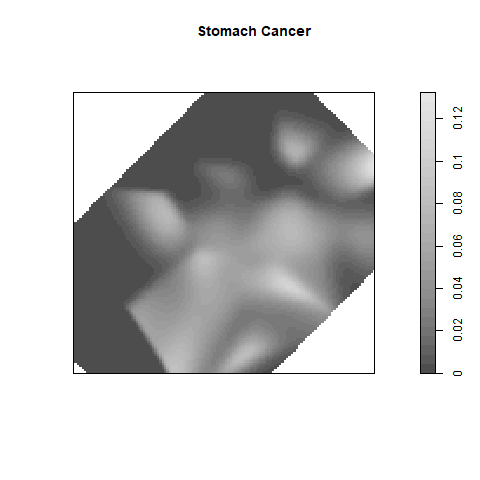}
\includegraphics[width=3.75cm]{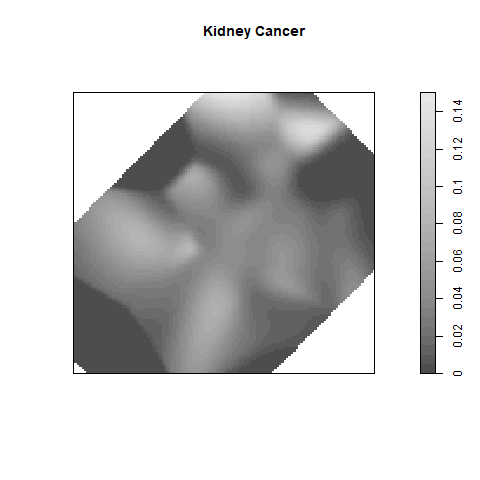}
\caption{Estimated intensities of the controls $\hat\lambda_0(x)$ and the cases $\hat\lambda_i(x)=1,2,3$ (top
row) and estimate of the ratio $\hat\rho_i(x),i=1,2,3$ (bottom row).}
\label{fig:intensities}
\end{figure}

Figure~\ref{fig:intensities} shows the estimates of the intensities for the
three types of cancer  $\hat\lambda_i(x)=1,2,3$ using a kernel smoothing with
a bandwidth of 300 meters, together with the estimates of the
ratio of the intensities $\hat\rho_i(x),i=1,2,3$. Note that, because of the
different number of points, the intensity and relative risk estimates are in
different scales.

The plots in Figure~\ref{fig:intensities} are only presented as a summary of
the spatial distribution of the point patterns, and the relative risk of the
different types of cancer.  A Monte Carlo test could be employed to find
regions of significant high risk\cite{KelsallDiggle:1995SiM}. Finally, a
semi-parametric estimate of the intensities of the cases could be used in order
to assess the impact of the covariates (and the locations of the polluting
industries) in the spatial variation of the risk\cite{diggle2007differences}.
These important issues will be addressed later within a Bayesian framework to
develop a log-Gaussian Cox model which can account for the effect of the
covariates and estimate the residual spatial variation of the risk.

\section{Multivariate point patterns for disease mapping}
\label{sec:mpp}

Diggle et al. (2005)\cite{Diggleetal:2005} develop a suitable framework
for the analysis of multivariate point patterns for the study of different
strains of bovine tuberculosis. Each strain $i=1,\ldots,K$ is modeled using
a different intensity $\lambda_i(x)$, and the probability of being a 
case of a strain of type $i$ at location $x$ is given by:

$$
p_i(x) = \frac{\lambda_i(x)}{\sum_{j=1}^K \lambda_j(x)}
$$
\noindent

Our problem differs from the previous approach because, for case-control data,
the intensity of the controls $\lambda_0(x)$ will act as a baseline in order
to compare the intensity of the different types of diseases considered
$\lambda_i(x),\ i=1,\ldots,K$. As mentioned above, for disease $i$, a relative
risk could be computed as the ratio $\rho_i (x) = \lambda_i(x) / \lambda_0(x)$
and departures from the value $n_i/n_0$ will indicate significant spatial risk
variation.  Hence, this ratio can be used to describe spatial risk variation.
Most importantly, it can help in the detection of hotspots by identifying
regions of unusual high relative risk and assessing increased risk by exposure
to pollution sources.

\subsection{Spatial risk variation}

As described in previous sections, spatial models for multivariate point
patterns in Public Health should account for the spatial distribution of the
population and address estimation of the effect of possible risk factors,
putative pollution sources and any residual spatial variation not explained by
previous factors.

First of all, our methodological proposal starts with a simple model to
estimate the different intensities. In particular, the model proposed 
at a point $x$ of the study domain $\mathcal{D}$ is

$$ \log(\lambda_0(x)) = \alpha_0 + S_0(x); x\in\mathcal{D}
$$

$$
\log(\lambda_i(x)) = \alpha_i + S_0(x) + S_i(x),\ i=1,\ldots,K; x\in\mathcal{D}
$$
\noindent
where $\alpha_i$ is an intercept and $S_i(x)$ is a spatial Gaussian process
with a Mat\'ern covariance and $K$ is the number of diseases in the study. In
general, the role of $\alpha_i$ is to account for the number of observed points
and term $S_0(x)$ estimates the underlying spatial variation. Bayesian point
estimates (e.g., the posterior means) of the intensities estimated by these
models can be similar to the estimate obtained by kernel smoothing with the
appropriate bandwidth\cite{TornadosVirgilio}.

Note that the previous model is in fact a joint model and that $S_0(x)$ is
estimated using cases and controls.  Furthermore, it holds that 

\begin{equation}
\log(\lambda_i(x)) - \log(\lambda_0(x)) = \alpha_i - \alpha_0 + S_i(x),\ i=1,2,3; x\in\mathcal{D}
\label{eq:intensity}
\end{equation}

Hence, spatial effects $S_i(x), i=1,2,3$ measure any disease-specific residual
spatial variation not accounted for by the distribution of the controls. As
$S_i(x)$ measures departure from the spatial distribution of the controls, it
can be used to detect areas of high risk by inspecting the credible intervals
of its posterior distribution.

G\'omez-Rubio et al. (2015)\cite{TornadosVirgilio}  propose a similar model in
the context of a study on the spatial distribution of tornados according to
their increasing strength (from 0 to 5) in the Contiguous United States. They
consider 6 different types or tornados, with the mildest tornados providing a
baseline intensity (i.e., tornados with intensity 0 acted as 'controls').  They
plugged-in posterior estimates of the intensity of the tornados with intensity
0 as covariates to estimate the intensities of the other types of tornados.
However, this ignores the uncertainty about the plugged-in intensity and the
model proposed in this paper should be preferred to model multivariate
point patterns within a Bayesian framework.

\subsection{Exposure to pollution sources}

If part of the spatial variation of the cases is thought to be 
explained by exposure to any of $P$ risk factors and
the spatial distribution of the population, its intensity could
be expressed as\cite{diggle2007differences}:

\begin{equation}
\lambda_i(x) = \lambda_0(x)\exp\{\alpha_i + \mathcal{F}_{ij}(x) + S_i(x)\}, i=1,\ldots,K; j=1\ldots,P
\label{eq:logCox}
\end{equation}
Here, $\mathcal{F}_{ij}(x)$ is a generic term that represents exposure to a
risk factor $j$ of a case of disease $i$, with $j=1,\ldots, P$.  For exposure
to pollution sources, this term usually depends on the distance to the location
of the pollution source.  The distance to the source for subject at location
$x\in\mathcal{D}$ to the pollution source will be denoted by $d_x$. Note that
this way of modeling exposure can be used without loss of generality to
represent other risk factors, such as exposure to pollutants, temperature, body
mass index, etc. and that it does not necessarily needs to be a distance.

Effect $\mathcal{F}_{ij}(x)$ can take several forms. First of all, a fixed
effect can be considered, i.e., $\mathcal{F}_{ij}(x) = \beta_{ij} d_x$.  This
means that exposure is modeled as a fixed effect in the linear term and that
each pollution source (or risk factor) $j$ affects differently each disease
$i$.  Negative values of $\beta_{ij}$ will indicate that the effect of the
pollution source is to increase the intensity of the point pattern  in its
vicinity.

However, this is seldom a good idea as usually modeling risk factors requires
a smoother term\cite{Wood:2017}. For this reason, two other smooth models will be
considered. The first one is a discrete random walk
based on $r$ knots placed at distances $k_1, \ldots, k_r$ from the pollution
source. For a given disease $i$, the random walk associated to pollution source
 $j$ is defined as 

$$
u_{lj} - u_{l-1j} \sim N(0, \tau_j); l = 2, \ldots, r
$$
with $\tau_j$ being a precision parameter. The effect is then defined as

$$
F_{ij}(x) = u_{i(x)j}.
$$
\noindent
Here, $i(x)$ is an index that indicates the nearest knot to the case
at location $x$. Hence, cases with similar distance to the pollution
source will be allocated the effect of the same random walk term and their
exposure will be similar.

The third effect will be a one-dimensional Gaussian process with a
Mat\'ern covariance based on the distance to the pollution source $j$.
This is similar to the one used to model spatial effects $S_i(x)$ but
using a single dimension.

In this case, the effect can be represented as

$$
F_{ij}(x) = v_{ij}(x)
$$
where $v_{ij}(x)$ is defined using a Gaussian process with zero mean and
variance defined by a Mat\'ern covariance in one dimension, where the values of
the observations are the distances from the points to the pollution source.

Note that all these three possible ways of modeling the exposure to the
pollution source can be extended to consider other risk factors which are
measured using continuous variables such as, for example, blood pressure,
weight, body-mass index, etc. Categorical risk factors, such as gender, can
also be considered using fixed effects. Furthermore, note that different
effects for different diseases are considered in our model, as exposure will
likely have different effects on different diseases.

%Finally, conditioning on the locations of cases and controls, the probability
%of being a case becomes\cite{diggle2007differences}:
%
%$$
%p_i(x) =\frac{\lambda_i(x)}{\lambda_0(x)+\lambda_i(x)}\
%$$
%\noindent
%and the logit of the previous probability is:
%
%$$
%logit(p_i(x)) = \log\left(\frac{\lambda_i(x)}{\lambda_0(x)}\right)= 
%\alpha_i - \alpha_0 + \mathcal{F}_{ij}(x) + S_i(x)
%$$

\subsection{Detection of regions of high risk}

Table~\ref{tab:ftdmodels} summaries  four variations of the proposed
model that can be fit to the data.
Model 0 can be regarded as a baseline model for comparison as it assumes that
the distribution of the cases and the controls is the same, and the intensities
are scaled according to the values of $\alpha_i,i=0, \ldots,K.$

Model 1 includes a disease-specific term $S_i(x)$ that accounts for unexplained
spatial variation. If the distribution of the cases is exactly that of the
controls, then this term should be equal to zero at every point of the study
region. For this reason, the posterior distribution of $S_i(x)$ can be
inspected for significant departures from zero, that will indicate
regions of unexplained high risk. For example, credible intervals
can be computed at grid points and then assess whether zero is inside.
Alternatively, the posterior mean could be computed or the posterior
probabilities of $S_i(x)$ being higher than zero could be computed as well (but
this is computationally more expensive).

Model 2 assumes that all spatial variation of the cases is explained by that of
the controls plus the effect of the risk factors. Model 3 adds a
disease-specific spatial term to account for unexplained spatial variation,
which can be used to assess regions of high risk not explained by the risk
factors.

Ideally, model selection criteria should help to decide which model is
best to explain the data. If either Model 1 (or Model 3) are selected
it means that there is some residual variation not explained by
the spatial distribution of the controls (and the risk factors)

\begin{table}
\begin{tabular}{ccll}
\hline
Model & $\log(\lambda_0(x))$ & $\log(\lambda_i(x))$\\
\hline
0 & $\alpha_0 + S_0(x)$ &  $\alpha_i + S_0(x)$\\
1 & $\alpha_0 + S_0(x)$ &  $\alpha_i + S_0(x) + S_i(x)$\\
2 & $\alpha_0 + S_0(x)$ &  $\alpha_i + S_0(x) + \mathcal{F}_{ij}(x)$\\
3 & $\alpha_0 + S_0(x)$ &  $\alpha_i + S_0(x)  + \mathcal{F}_{ij}(x) + S_i(x)$ \\
\hline
\end{tabular}
\caption{Summary of models for multivariate point patterns using case-control data.}
\label{tab:ftdmodels}
\end{table}

Furthermore, this joint modeling approach allows for the comparison of
different disease-specific spatial terms, i.e., $\Delta_{uv}(x) = S_u(x) -
S_v(x),\ u\neq v$, can be computed to assess similar residual spatial variation
between two types of cancer. In the case of spatial residual variation, values
of $\Delta_{uv}(x)$ close to zero will indicate a similar residual spatial
variation which may point to common risk factors between two
types of cancer.

\section{Integrated Nested Laplace Approximation}
\label{sec:inla}

Model fitting will be carried out using the integrated nested Laplace
approximation (INLA)\cite{INLA}. This approximation assumes that the model can
be expressed as a latent Gaussian Markov random field
(GMRF)\cite{RueHeld:2005}, i.e., the latent effects have Gaussian distribution
with sparse covariance matrix that may depend on further hyperparameters.  This
includes some widely used models, such as, generalized linear and additive
models with random effects. INLA will only provide approximations to the
marginal posterior distributions of the model effects and hyperparameters, but
this is often enough for inference.

INLA can fit models with latent spatial terms $S_i(x),i=0,\ldots,K$ that are
defined using a Mat\'ern covariance using the approximation based on stochastic
partial differential equations (SPDE)\cite{SPDE,Krainskietal:2019}.  Furthermore, the
multivariate point patterns model will be implemented using a representation of
the cases and the controls as a Poisson process\cite{SPDELog-GausianCox}.

The Mat\'ern covariance between two points $x_1$ and $x_2$, separated by a
distance $d_{12}$, is defined as

$$
Cov(x_1, x_2) = \sigma^2 
\frac{2^{1-\nu}}{\Gamma(\nu)}
(\kappa d_{12})^\nu
K_\nu(\kappa d_{12})
$$
Here, $\sigma^2$ is a variance parameter, $\kappa$ a spatial scale parameter
and $\nu$ a smoothness parameter. Furthermore, $\Gamma(\cdot)$ is the Gamma
function and $K_\nu(\cdot)$ is the modified Bessel function of the second kind.

The SPDE approximation\cite{SPDE} to estimate a Gaussian process with Mat\'ern
covariance $S(x)$ implemented is based on a weak solution to a SPDE, so that
the estimated effect is expressed as

$$
S(x) = \sum_{k=1}^m \psi_k(x) w_k
$$
Here, $\psi_k(x)$ are a basis of functions and $w_k$ are Gaussian weights.
$k$ represents the number of vertices in a triangulation that covers the
study region. This triangulation is used to define the basis functions
as each function is piecewise linear within each triangle. In particular,
$\psi_k(x)$ is equal to 1 at vertex $k$ and 0 at all other vertices. This
provides a sparse representation that is very convenient in practice
for computational purposes.
When reporting the results, the nominal variance and the nominal range, will be
used to summarize the estimates of the spatial effects $S_i(x),i=0,\ldots,K$.

Fitting log-Gaussian Cox processes to point patterns with INLA is based on
reformulating the model as a Poisson regression\cite{SPDELog-GausianCox}. This
requires creating a Voronoi tessellation using the observed points of the point
pattern.  For each polygon, a dummy observation with value 0 and associated
value $A_k$, the area of the associated polygon, is added, and each observed
point is added with a value 1 and $A_k$ equal to 0. Then, the model is a Poisson
regression as follows:

$$
 y_k \sim Po(A_k \theta_k), k = 1, \ldots
$$
\noindent
Here, $y_k$ are the 0/1 values described above, $A_k$ the area of the
associated polygon and $\theta_k$ accounts for the  model effects. For example,
for Model 3 and disease $i$ this would be:

$$
\log(\theta_k) = \alpha_i + S_0(x_k) + \mathcal{F}_{ij}(x_k) + S_i(x_k)
$$
where $x_k$ is the location of observed or dummy point $x_k$.

Note that, in order to fit this model with INLA, the covariates associated to the risk
factors need to be
available at the dummy points. This is not a problem when they are distances to pollution sources, but it could be a problem when the risk factor is 
an individual level variable, such as gender or age.

In practice, implementing this joint model requires the use of different
likelihoods, one for each point pattern, with a shared effect $S_0(x)$.  This
is fully described in the R code used in this paper, which is 
available from Github at \url{https://github.com/becarioprecario/INLA_MVPP}.

%included in the Supplementary Material of this
%paper.

%\section{Example: Cancer risk in Alcal\'a de Henares (Madrid, Spain)}
%\section{Assessing spatial risk variation around pollution sources}
\section{Spatial variation of cancer in Alcal\'a de Henares (Madrid, Spain)}
\label{sec:example}

We have analyzed the data introduced at the beginning of the paper on cancer
data from Alcal\'a de Henares (Madrid, Spain) using the model introduced above.
In order to fully describe the model, the priors on the different parameters
must be stated. Priors on the fixed effect have been a Gaussian distribution
with zero mean and precision 1000. We have used PC-priors
\citep{Simpsonetal:2017} for the parameters of the SPDE-based spatial effects.
In particular,  the prior on the range is so that $P(range < 5) = 0.95$ and the
prior on the standard deviation $\sigma$ is so that $P(\sigma > 10) = 0.01$.
These setting are based on reasonable vague assumptions about the underlying
spatial processes.

\subsection{Confounding factors}

When assessing increased risk around pollution sources it is important to
account for other factors that may play a role in the spatial
distribution of the disease. For example, socioeconomic indicators
(e.g., income or education) or life style (e.g., diet or smoking status)
may drive the spatial pattern for some diseases. For example, lung cancer
is highly correlated with income and smoking\citep{Faggianoetal:1997,Spitzetal:2006}.

For this reason, the models presented here can include fixed effects on some
covariates. Note that given the multivariate nature of our models it is
important to allow for disease-specific coefficients for the fixed effects
because different risk factors can affect the diseases under study in different
ways.

Furthermore, our analysis will consider models that do not account for risk
factors first. Then, models that account for confounding factors will be fit to
assess whether the effects from pollution sources are still significant.  This
will allow us to determine whether the primary cause of increased mortality is
due to the pollution source or socio-economic confounding variables.

In particular, the variables that we have considered are available at the
census tract level, obtained from the 2001 Spanish census by Instituto Nacional
de Estad\'istica (INE, Spain). Values have been assigned to the cases or the
controls by matching the census tract. The variables considered are
unemployment rate in the range between 20 and 59 years (\textit{UNEMP2059}), an
average of the score
for social class (\textit{AVGSOC}), an average of the score for education
level in the range between 30 and 39 (\textit{AVGEDU}) and the percentage of
children aged between 0 and 3 that are in the school (\textit{PCTSCH}).

\subsection{Spatial risk variation}

The first step in our analysis of the three types of cancer in Alcal\'a de
Henares will be to inspect spatial risk variation in order to assess whether
the spatial distributions of the cancers is the same as the controls. For this,
we have fit models 0 and 1, and we have computed some model selection
measures in order to make a decision on the best model. Models 2 and 3 are
discussed later as they will depend on the pollution sources included in the
model.

\begin{table}
\centering
\begin{footnotesize}
\begin{tabular}{|c|c|c|c|c|c|c|}
\hline
 & \multicolumn{3}{|c|}{No confunding} & \multicolumn{3}{|c|}{With confunding}\\
 & \multicolumn{3}{|c|}{factors} & \multicolumn{3}{|c|}{factors}\\
\hline
Model & DIC & WAIC & Marg. lik. & DIC & WAIC & Marg. lik.\\
%& \multicolumn{3}{c|}{$S_0(x)$} & \multicolumn{3}{c|}{$S_1(x)$} \\
\hline
0 & -31454.12 & -28463.91 & 15558.89 & -31441.31 & -28390.93 & 15494.42\\
1 & -31486.21 & -28365.77 & 15560.84 & -31479.50 & -28263.28 & 15498.55\\
\hline
\end{tabular}

\end{footnotesize}
\caption{Model selection criteria values for the different models fit to the case-control data.}
\label{tab:models}
\end{table}

Table~\ref{tab:models} summarizes the different criteria computed with INLA.
For models without confounding factors, the DIC and marginal likelihood support Model 1, i.e.,
that there is disease-specific spatial variation not explained by the spatial
distribution of the controls.  Models that include confounding factors do not
seem to improve model fitting. Table~\ref{tab:conffactors} shows the effects of
the covariates. In particular, high unemployment and children in school are
close to have a significant positive association with the three types of
cancer. On the other hand, education level and social level does not seem to
have an effect.

These four covariates where taken not to be correlated with each other but it
is possible that some confounding is occurring when all four are in the model,
and this might be the reason why all credible intervals contain the zero value
and model selection criteria do not favor models with confounding factors
included. In any case, by including these four covariates typical possible
socio-economic confounders are taken into account and we have kept all four
variables in all models that include confounding factors.

%Note also that Model 1 with confounding factors seems to have an
%identification issue as the estimates that we have obtained do not look very
%reasonable
%(see below). This may happen because the confounding factors explain most of
%the disease-specific spatial residual variation. For this reason, when
%confounding factors are included, we will only consider models without
%disease-specific spatial effects (i.e., Models 0 and 2).

%PIT also
%support this claim. On the other hand, WAIC and CPO, seem to prefer Model 0,
%with the marginal likelihood providing similar value for both models.

\begin{table}
\centering
\begin{footnotesize}
\begin{tabular}{|c|c|rr|rr||rr|rr|}
\hline
 Variable & Cancer & \multicolumn{4}{|c||}{Model 0} & \multicolumn{4}{|c|}{Model 1}\\
\hline
 & & Mean & St. dev. & \multicolumn{2}{|c||}{95\% C.I.} & Mean & St. dev. & \multicolumn{2}{|c|}{95\% C.I.}\\
\hline
UNEMP2059 & lung & 0.03 & 0.03 & -0.02 & 0.09 & 0.06 & 0.03 & -0.00 & 0.13\\
  UNEMP2059 & stomach & 0.05 & 0.04 & -0.03 & 0.13 & 0.09 & 0.05 & -0.01 & 0.18\\
  UNEMP2059 & cancer & 0.06 & 0.05 & -0.03 & 0.15 & 0.08 & 0.05 & -0.02 & 0.18\\
  AVGSOC & lung & 0.80 & 1.07 & -1.30 & 2.91 & 1.81 & 1.19 & -0.51 & 4.16\\
  AVGSOC & stomach & -0.30 & 1.58 & -3.38 & 2.81 & 0.86 & 1.75 & -2.55 & 4.33\\
  AVGSOC & cancer & 0.01 & 1.67 & -3.25 & 3.32 & 0.62 & 1.83 & -2.90 & 4.27\\
  AVGEDU & lung & -0.11 & 0.35 & -0.79 & 0.57 & -0.21 & 0.40 & -0.99 & 0.58\\
  AVGEDU & stomach & 0.21 & 0.50 & -0.78 & 1.20 & -0.10 & 0.57 & -1.23 & 1.01\\
  AVGEDU & cancer & 0.47 & 0.55 & -0.61 & 1.53 & 0.31 & 0.60 & -0.90 & 1.47\\
  PCTSCH & lung & 0.01 & 0.01 & -0.00 & 0.02 & 0.01 & 0.01 & -0.00 & 0.3\\
  PCTSCH & stomach & 0.00 & 0.01 & -0.01 & 0.02 & 0.01 & 0.01 & -0.01 & 0.03\\
  PCTSCH & cancer & 0.01 & 0.01 & -0.01 & 0.03 & 0.02 & 0.01 & -0.01 & 0.04\\
\hline
\end{tabular}

\end{footnotesize}
\caption{Estimates of the confounding factors included in the models
for Models 0 and 1.}
\label{tab:conffactors}
\end{table}

Table~\ref{tab:effects} summarizes the spatial effects $S_i(x)$ of the
different models. This table is useful to compare the estimates of effect
$S_0(x)$ between the different models.  When no confounding factors are
included, we find very similar estimates of the parameters of $S_0(x)$ in both
models.  Under Model 1 stomach cancer seems to have the highest nominal
standard deviation and nominal range of the disease-specific spatial effect,
which may point to a differential spatial variation. On the other hand, kidney
cancer has the smallest nominal variance and range, which may point to a lack
of differential spatial distribution. Finally, the estimates of the parameters
for the spatial effect of lung cancer point to a possible mild differential
spatial variation.

When confounding factors are included, the spatial effect in Model 0 has very
similar estimates as the case with no confounding factors. However, Model 1
shows smaller estimates of the range for the disease-specific spatial effects.
In our opinion, this is due to the socio-economic confounding variables included
in the model, that account for some of the spatial variation in the data.

\begin{table}
\centering
\begin{footnotesize}
\begin{tabular}{|c|c|c|cc|cc|}
\hline
Model & Cancer  & Conf. factor & \multicolumn{2}{c|}{Nominal Range} & \multicolumn{2}{c|}{Nominal St. dev.} \\ \hline
      &         & &  Mean &    St. dev. & Mean & St. dev. \\
\hline
0    & Controls & No & 3.94 & 1.13 & 4.06 & 1.08 \\
1    & Controls & No & 3.89 & 1.16 & 3.99 & 1.10 \\
1    & Lung     & No & 6.64 & 6.52 & 0.76 & 0.44 \\
1    & Stomach  & No & 6.91 & 6.08 & 0.85 & 0.51\\
1    & Kidney   & No & 3.00 & 8.76 & 0.22 & 0.15\\
\hline
\hline
0    & Controls & Yes & 3.93 & 1.12 & 4.04 & 1.07\\
1    & Controls & Yes & 3.88 & 1.09 & 4.00 & 1.04\\
1    & Lung     & Yes & 4.00 & 2.70 & 0.60 & 0.22\\
1    & Stomach  & Yes & 5.56 & 4.40 & 0.86 & 0.38\\
1    & Kidney   & Yes & 1.45 & 1.64 & 0.33 & 0.17\\
\hline
\end{tabular}

\end{footnotesize}
\caption{Summary of the spatial effects of the different models fit to the case-control data.}
\label{tab:effects}
\end{table}

Because of this small sample size, detecting departures from Model 0 will be
difficult. Given the nature of the problem at hand, it is not easy to increase
the sample size, and the only reason to do this is to extend the period of time
of the analysis, which is not always possible. A similar problem will be faced
when assessing exposure to pollution sources. For this reason, a simulation
study will be developed at the end of this section.

\subsection{Detection of regions of high risk}

Given that the underlying spatial variation $S_0(x)$ is taken as a baseline as
it represents the distribution of the controls, the detection of the areas of
high risk can be tackled by looking at which points in the study region have
high (or low) values of $S_i(x)$ using its posterior distribution
$\pi(S_i(x)|y)$.  We will inspect this by looking at the credible intervals of
the estimates of the disease-specific effects computed at a grid of points
inside the study region. Departures from the null value will indicate regions
of high or low intensity.

\begin{figure}
\centering

\includegraphics[scale=0.25]{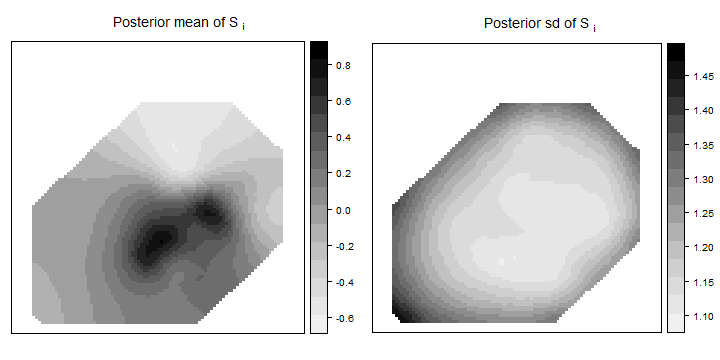}
\includegraphics[scale=0.22]{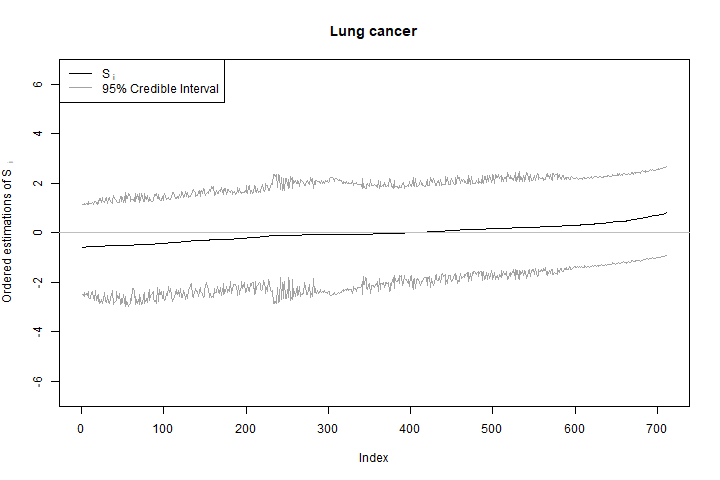}

\includegraphics[scale=0.25]{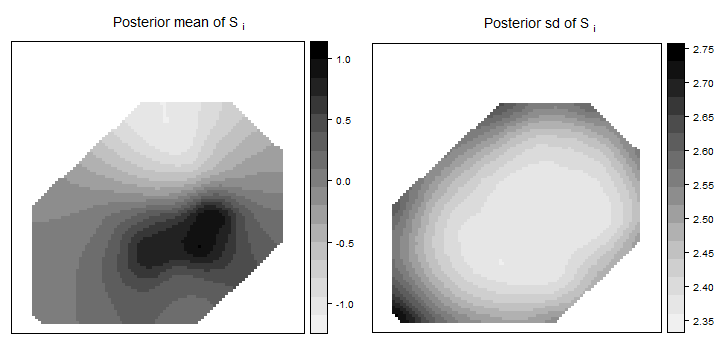}
\includegraphics[scale=0.22]{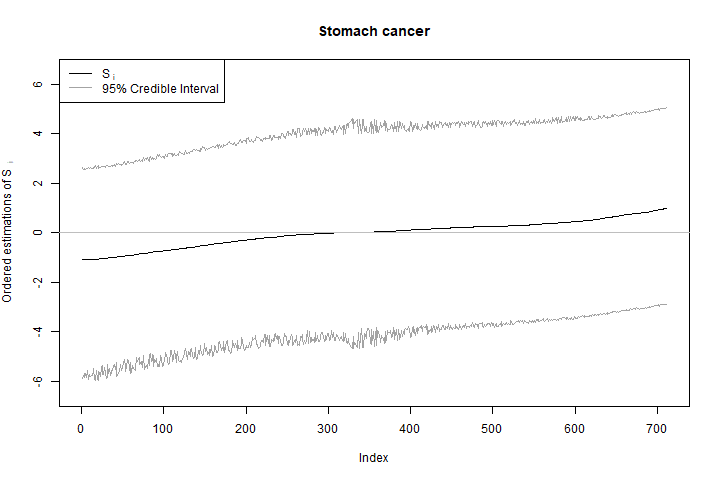}

\includegraphics[scale=0.25]{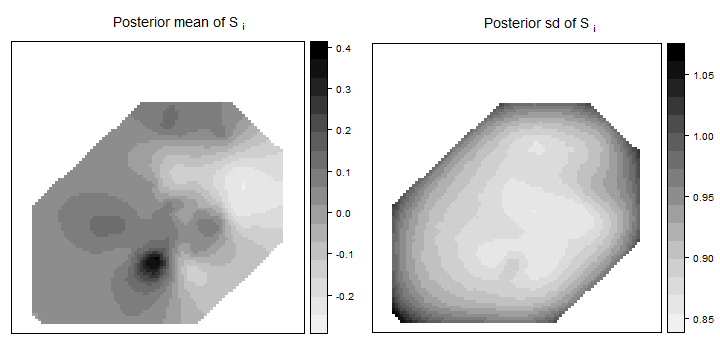}
\includegraphics[scale=0.22]{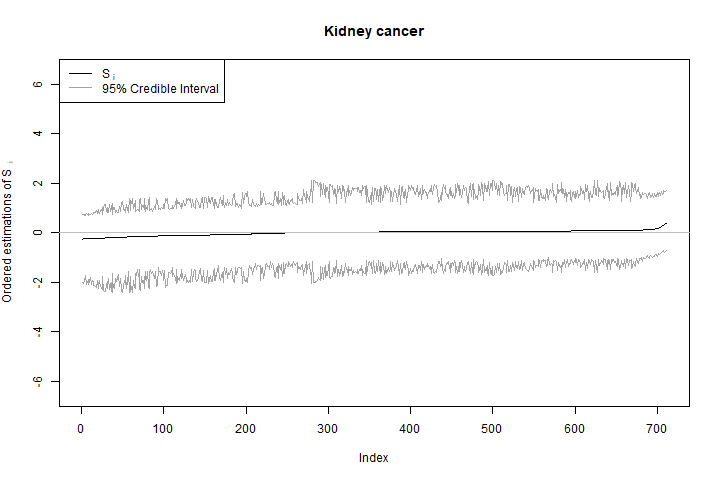}

\caption{Summary statistics of spatial effects $S_i(x)$:
map of posterior means (top row), posterior means and 95\% credible intervals
(middle row) and map of posterior probabilities $\pi(S_i(x)>0|y)$.}
\label{fig:SPDE1}
\end{figure}

Figure~\ref{fig:SPDE1} displays posterior means and standard deviations of the
spatial effects $S_i(x)$ for the different types of cancer obtained by fitting
Model 1 and accounting for confounding factors. In the plots, point estimates and
credible intervals have been arranged in increasing order (using the posterior
mean).  These intervals can be used to assess whether residual spatial
variation has a high probability of being different from zero, which will
indicate a departure in the cases from the spatial distribution of the
controls.

The estimates provided by Model 1 without adjusting for confounding factors are
similar, but with wider credible intervals (and they have not been included
here). The reduction of the width of the credible intervals is then due to the
effect of the covariates. Because of the small sample size, we believe that
credible intervals are not narrow enough as to detect hostspots due to the
polluting industries. For these reasons, we have decided to consider the models
that include exposure to pollution sources in the analysis.

For kidney cancers, all credible intervals contain the zero value, which
indicates no departure from the underlying spatial distribution of the
controls. However, lung and stomach cancers have higher point estimates of
$S_i(x)$ and they are close to having smaller regions with high values of the
disease-specific spatial term. This indicates spatial variation not accounted
for the spatial distribution of the controls. Note how these regions seem to be
close to the south part of the city. Hence, we believe that it is important to
test for a possible association between this increased intensity in the cases
and the location of industries. This is carried out below, using models 2 and
3.

%As stated above, Model 1 with confounding factors may be overparametrized
%according to the estimates of the parameters of the spatial effects.  Our
%results show that all diseases have now regions of high risk, but this may be
%an artifact of including covariates in the model. For this reason, we will not
%consider it here.

%"V48",#Tasa paro 20-59 
%  "V53", # COndición socioeconómica media
%  "V54", #"Nivel medio de estudios en el grupo de edad 30-39"
%  "V55" #Estudios pre-obligatorios (%)" 

\subsection{Assessing exposure to pollution sources}

We will inspect a possible association between the location of the industries
in the city and an increased intensity about them.  Full details about the
estimates for the different models are included in the Supplementary Materials
of this paper. In the analysis of these results, we will focus on the models
that point to an increased risk around polluting industries. For this reason,
model selection will not only be based on the DIC and WAIC criteria but also on
reasonable estimates of the spatial effects in the model and on estimates of
the effect $\mathcal{F}_{ij}(x)$ that points to an increased risk around the
pollution source. Note that some estimates of the DIC and WAIC were not
reliable and they have been replaced by a dash in some of the tables in this
paper.

\begin{figure}
\centering
%\includegraphics[scale=0.5,trim={0.5cm 4.5cm 11cm 4.5cm},clip]{M2I5RW1NotS.pdf}
%\includegraphics[scale=0.5,trim={0.5cm 4.5cm 11cm 4.5cm},clip]{M2I5SPDE1.pdf}
%\caption{Estimated effects of Model 2 for the effect of the proximity to industry 5 using RW1 (left) and  SPDE1 (right) smooth effects.}
\includegraphics[width=12cm]{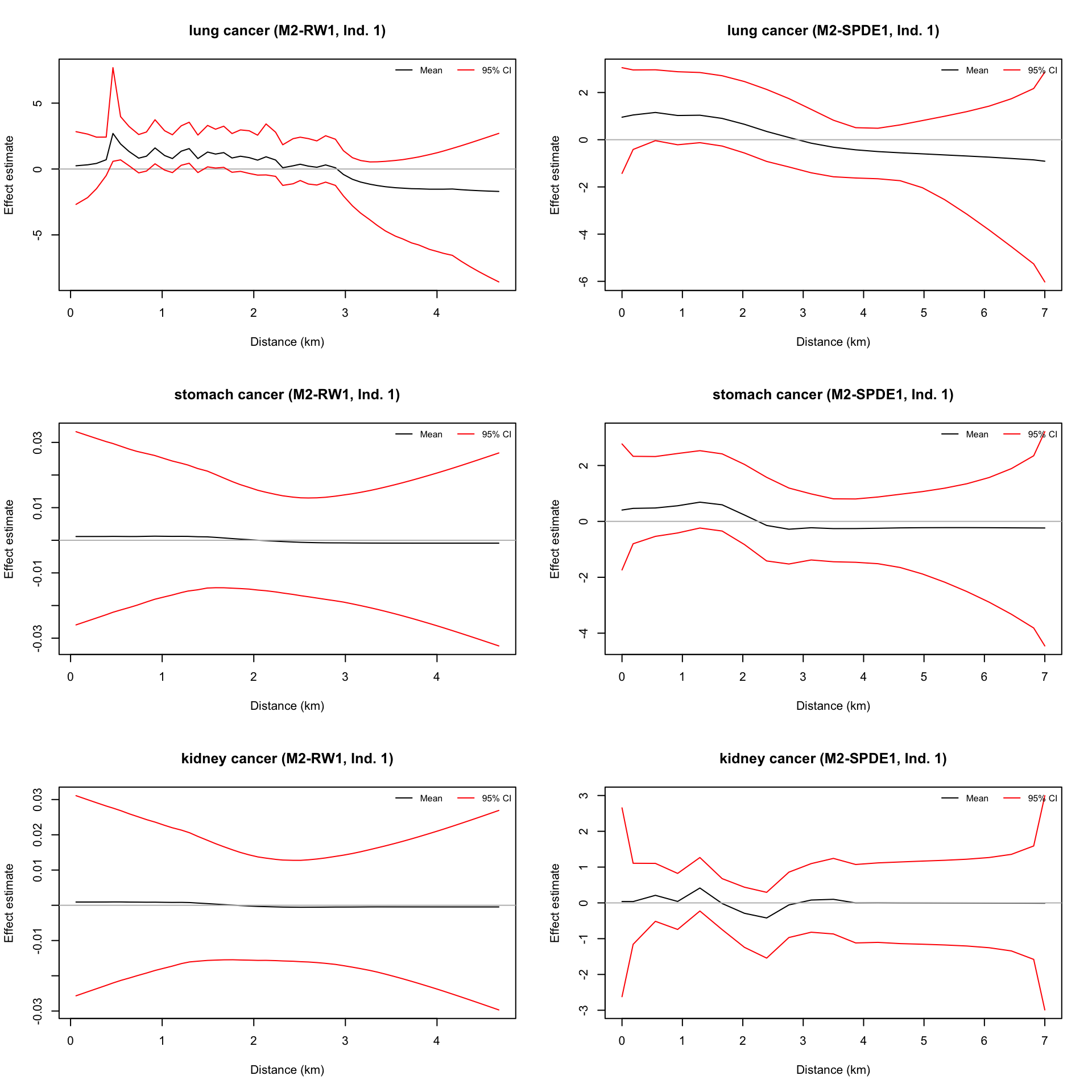}
\caption{Estimated effects of Model 2 for the effect of the proximity to industry 1 using RW1 (left) and  SPDE1 (right) smooth effects.}
\label{fig:smooth}
\end{figure}

Table~\ref{tab:Models23} shows estimates of the DIC and WAIC for the different
models (not adjusted for confounding variables) to assess risk around pollution
sources using a fixed effect, a RW1 and a SPDE1. This table also summarizes for
which diseases there is an increased risk. The inclusion criterion has been a
negative upper limit of the 95\% credible interval of the coefficient for fixed
effect, or a decreasing trend with distance when the effect is either a RW1 or
SPDE1 smooth term (even if the 95\% credible intervals contained the zero
value). The top plots in Figure~\ref{fig:smooth} show an example of the
inclusion criteria used to list a tumor in Table~\ref{tab:Models23}. Note
that the descending trend with distance is clear, but that for the RW1 and
SPDE1 effects 95\% credible
intervals contain the zero value. However, we believe this is simply due to the
small sample size that we have in our particular dataset.

In addition to the previous criteria, values of the DIC and WAIC smaller than
the ones obtained for Model 0 and 1 will indicate a significant effect. This
is associated with a credible interval for the fixed effect that is above zero,
and effects RW1 and SPDE1 showing a decreasing pattern around the pollution
sources, such as the one seen in Figure~\ref{fig:smooth} (for Industry 1).

We have not reported here the estimates of the different spatial effects in the
model because, in general, these are very close to the obtained for Models 0
and 1. In any case, these are provided in the Supplementary Materials of this
paper.  The results indicate that seven industries have a potential association
with an increase of the intensity of different types of cancer around them.
This is particularly clear for lung and stomach cancer, while this possible
association with kidney cancer is very mild or inexistent.

%Figure~\ref{fig:smooth} shows the estimates of the RW1 and SPDE1 effect for
%Industry 1 using Model 2 and adjusting for confounding factors (without
%adjusting for them the results are very similar). As it can be seen,
%there is a clear decrease of the effect (which also means a decrease in the
%intensity of the cases as the distance to the pollution source increases) for
%lung and stomach cancer. This effect cannot be appreciated for kidney cancer.
%Note that 95\% credible intervals  have been plotted and that they contain the
%value zero, but this may be due to the small sample size of the cases. In order
%to show that these methods can detect effects, provided a sufficient sample
%size, a simulation study has been included later.

Given that this increase may also be due to socio-economic factors, we have fit
the same models with the four socio-economic variables mentioned above.  In
general, the estimates of the coefficients are very similar to those obtained
for Models 0 and 1, and they are not reported here (but they have been included
in the Supplementary Materials). Table~\ref{tab:Model2conf} shows a summary of
the DIC and WAIC for these models, that includes for which types of cancer
there is a significant increase around the pollution source.  As mentioned
earlier, Model 3 has not been included here because we suspect that the
different effects in the model are not identifiable as we are accounting for
confounding factors and including disease-specific spatial terms.

Values of the DIC are smaller when adjusting for confounding factors. However,
WAIC seems to increase. When assessing exposure, models with RW1 and SPDE1 are
preferred over models with fixed effects. In general, the associations detected
by the different models are very similar to the case with no adjustment, which
means that there is still a possible association between an increase in the
number of cases and the distance to the polluting industries.  This association
is clear for lung and stomach cancer, and very mild (or inexistent) for kidney
cancer.

Regarding the industries that appear in Table~\ref{tab:Models23} and
Table~\ref{tab:Model2conf}, Industries 1 and 2 are part of an industrial area
very close to the city center. The presence of asbestos in this area could
explain the apparent increase in the cases of lung and stomach cancer. Industry
5 is a landfill, where waste is often cremated and the smoke reaches the
population at kilometers away. Industry 6 is close to a deprived area, which
could be the case of this increase in the cases of cancer but we have already
accounted for several socioeconomic variables. Industries 7, 8 and 9 do not
show any association for the models with RW1 and SPDE1 effects and we believe
that there is in fact no association with cancer.

Given that our study is limited by the small sample size of the cases and the
four socio-economic variables included in the model, we want to be cautious
about pointing to any significant association between the increase of cases of
cancer around the aforementioned industries. However, we believe that the new
methodology developed in this paper is appropriate for the task at hand.

\begin{table}
\centering
\scriptsize
\begin{tabular}{|c|c|ccc||ccc|}
\hline
 & & \multicolumn{3}{|c||}{Model 2} & \multicolumn{3}{|c|}{Model 3}\\
\hline
Source & Effect & DIC & WAIC & Tumour & DIC & WAIC & Tumour\\
\hline
Industry 1 & Fixed & -31576.57 & -28523.96 & L, S & -31604.58 & -28446.86 & L, S\\
Industry 2 & Fixed & -31575.79 & -28522.54 & L, S & -31605.73 & -28444.19 & L, S\\
Industry 5 & Fixed & -31576.32 & -28525.60 & L, S & -31598.35 & -28432.65 & --\\
Industry 6 & Fixed & -31574.03 & -28516.32 & L, S, K&-31603.99 & -28440.13 & L\\
Industry 7 & Fixed &-31561.54 & -28501.01 & -- & -31605.74 & -28438.69 & L\\
Industry 8 & Fixed & -31565.95 & -28509.09 & L, K &-31599.19 & -28435.79 & --\\
Industry 9 & Fixed & -31566.34 & -28509.38 & L, K & -31599.81 & -28436.22 & --\\
\hline
\hline
Industry 1 & RW1 & -31563.00 & -28491.15 & L & -31604.58 & -28446.86 & L, S\\
Industry 2 & RW1 & -31621.04 & -27630.06 & L & -31605.73 & -28444.19 & L, S\\
Industry 5 & RW1 & -31618.46 & -28490.14 & L, S & -31598.35 & -28432.65 & L, S\\
Industry 6 & RW1 & -31557.60 & -28497.06 & L & -31603.99 & -28440.13 & L, S\\
Industry 7 & RW1 & -31559.55 & -28519.13 & -- & -31605.74 & -28438.69 & --\\
Industry 8 & RW1 & -31559.63 & -28515.80 & --& -31599.19 & -28435.79 & --\\
Industry 9 & RW1 & -31559.61 & -28515.22 & -- & -31599.81 & -28436.22 & --\\
\hline
\hline
Industry 1 & SPDE1 & -28613.62 & -- & L, S& -- & -- & L, S\\
Industry 2 & SPDE1 & -31573.44 & -28447.03 & L, S & -- & -- & L, S\\
Industry 5 & SPDE1 & -31606.01 & -- & L, S& -31590.59 & -- & L, S\\
Industry 6 & SPDE1 & -31590.62 & -28438.07 & L& -31604.02 & -28376.68 & --\\
Industry 7 & SPDE1 & -31568.83 & -28451.21 & ---& -31600.12 & -28398.24 & --\\
Industry 8 & SPDE1 & -31567.08 & -28448.97 & ---& -31594.16 & -28384.54 & --\\
Industry 9 & SPDE1 & -31565.64 & -28454.97 & ---& -31593.62 & -28294.52 & --\\
\hline
\end{tabular}

\caption{Summary of Models 2 and 3 (without confounding factors) to assess
exposure around polluting industries.}
\label{tab:Models23}
\end{table}

\begin{table}
\centering
\scriptsize
\begin{tabular}{|c|c|ccc|}
\hline
 & & \multicolumn{3}{|c|}{Model 2}\\
\hline
Source & Effect & DIC & WAIC & Tumour \\
\hline
Industry 1 & Fixed & -31624.75 & -28461.34 & L, S, K\\
Industry 2 & Fixed & -31623.27 & -28461.01 & L, S, K\\ 
Industry 5 & Fixed & -31598.99 & -28449.47 & L, S, K\\
Industry 6 & Fixed & -31609.41 & -28451.30 & L, S, K\\
Industry 7 & Fixed & -31585.11 & -28429.39 & K\\
Industry 8 & Fixed & -31600.10 & -28445.24 & L, K\\
Industry 9 & Fixed & -31601.54 & -28445.97 & L, K\\
\hline
\hline
Industry 1 & RW1 & -31645.50 & -- & L, S\\
Industry 2 & RW1 & -31734.66 & -- & L, K\\
Industry 5 & RW1 & -31657.11 & -28415.90 & L, S\\
Industry 6 & RW1 & -31773.61 & -- & L, S\\
Industry 7 & RW1 & -31583.83 & -28451.82 & --\\
Industry 8 & RW1 & -31613.32 & -28388.08 & L\\
Industry 9 & RW1 & -31609.66 & -28381.25 & L\\
\hline
\hline
Industry 1 & SPDE1 & -31626.01 & -- & L, S\\
Industry 2 & SPDE1 & -31632.65 & -- & L, S\\
Industry 5 & SPDE1 & -31661.90 & -- & L, S\\
Industry 6 & SPDE1 & -31682.90 & -28289.75 & --\\
Industry 7 & SPDE1 & -31619.81 & -28375.44 & --\\
Industry 8 & SPDE1 & -31622.06 & -28370.71 & --\\
Industry 9 & SPDE1 & -31619.74 & -28353.05 & --\\
\hline
\end{tabular}

\caption{Summary of Model 2 (with confounding variables) to assess exposure around polluting industries.}
\label{tab:Model2conf}
\end{table}

Given that Model 2 seems to be the best model for RW1 and SPDE1 effects, we
have displayed the estimates of the effects for Industry 1 in Figure
\ref{fig:smooth}.  The estimated effects are similar between RW1 and SPDE1
effects, with a step-like effect for lung and stomach cancer.  This is
consistent with the fact that this pollution source is close to the city
center.  The effect on kidney cancer is negligible. Similar figures for all the
other pollution sources are available in the Supplementary Materials.

%\begin{figure}
%\centering
%\includegraphics[scale=0.65,trim={0.5cm 4.5cm 0.5cm 4.5cm},clip]{M2I5RW1NotS.pdf}
%\caption{Estimated effects of Model 3 for the effect of the proximity to industry 5 using a RW1 smooth effect.}
%\label{fig:M2I5RW1}
%\end{figure}
%
%
%
%\begin{figure}
%\centering
%\includegraphics[scale=0.65,trim={0.5cm 4.5cm 0.5cm 4.5cm},clip]{M2I5SPDE1.pdf}
%\caption{Estimated effects of Model 3 for the effect of the proximity to industry 5 using a SPDE1 smooth effect.}
%\label{fig:M2I5SPDE1}
%\end{figure}
%%

Detection of the effect produced by the pollution sources and other risk
factors depends on the sample size of the data. In epidemiological studies it
is seldom possible to increase the sample size of the cases and the association
between risk factors and the disease may not be detected because of a small
sample size. For this reason, we have conducted a simulation study in the next
section to assess how the estimates of the effects depend on the sample size of
the cases. In particular, we will pay attention to point estimates (i.e.,
posterior means) and 95\% credible intervals.

\subsection{Simulation study}

In order to assess how sample size impacts the detection of effects due to risk
factor using the the three forms of function $\mathcal{F}_{ij}(x)$, a
simulation study has been carried out. In particular, we have simulated a new
set of 3000 controls from the estimated intensity of the actual controls, and
then we have simulated a set of cases using the estimated intensity of the
simulated controls modulated by an effect that depends on the distance to a
pollution source using an exponential decay function. Note that cases from a
single disease are considered now instead of cases from several diseases.
Also, the putative pollution source used in the simulations is 
located where Industry 1 is (see Figure~\ref{fig:locations}).

This means that the intensity of the simulated cases is

$$
\hat\lambda_0(x) \exp\{-d_x/\phi\} 
$$
where $\hat\lambda_0(x)$ is the estimated intensity of the simulated controls,
$d_x$ the distance (in kilometers) to the pollution source and $\phi$ a scale
parameter are parameters to measure how the intensity of the cases is modulated
by the distance to the pollution source.

Data have been simulated using values of $\phi$ of 1, 3 and 6, and the number
of simulated cases have been 50, 100, 300, 500, 1000 and 2000.  The number of
simulated controls has always been 3000. Distances from the simulated points to
the pollution source range from 0.05 to 5.3, approximately. Hence, for a value
of $\phi$ equal to 1 we can expect a fast decay,
while a value of 6 will produce a slow decay of the effect of the pollution
source (and very similar effects of the source on the intensity of all the controls). 

Table~\ref{table:sim} shows the values of the DIC for Models 2 and 3 fit to the
simulated data using RW1 and SPDE1 effects on the distance to the pollution
source. RW1 models point to Model 2 in most cases and, in particular, for large
sample sizes and large values of $\phi$.  SPDE1 models show very similar values
for both models, which would lead to selecting Model 2 (as this is simpler).

In general, for $\phi$ equal to 6 both models
provide similar values of the DIC for both RW1 and SPDE1. This is consistent
with a situation in which the pollution source has a negligible effect.

Hence, detectability of the effect produced by the proximity to the pollution
source increases with sample size and the strength of the effect on the
proximity. For a conveniently large sample size, the models presented in this
paper are able to detect exposure to a pollution source, even when this effect
is mild.

\begin{table}
\centering
\small
\begin{tabular}{rl|ll|ll|ll}
  \hline
\multicolumn{2}{c|}{Settings} & \multicolumn{2}{c|}{} & \multicolumn{2}{c|}{RW1} & \multicolumn{2}{c}{SPDE1}\\
\# Cases & $\varphi$ & M0 & M1 & M2 & M3 & M2 & Model 3\\
  \hline
    50 & 1 & -22812.90 & -22831.77 & -22838.29 & -22831.98 & -22839.07 & -22839.12 \\ 
    50 & 3 & -22796.62 & -22796.39 & -22796.93 & -22796.62 & -22800.88 & -22800.81 \\ 
    50 & 6 & -22796.41 & -22796.12 & -22796.74 & -22796.15 & -22796.29 & -22795.99 \\ 
   100 & 1 & -22934.97 & -22965.97 & -22967.99 & -22965.95 & -22976.62 & -22976.27 \\ 
   100 & 3 & -22911.74 & -22913.33 & -22912.16 & -22912.62 & -22921.13 & -22920.72 \\ 
   100 & 6 & -22908.13 & -22907.91 & -22908.53 & -22908.13 & -22911.90 & -22911.18 \\ 
   300 & 1 & -23880.00 & -24028.13 & -24035.86 & -24028.87 & -24049.91 & -24049.47 \\ 
   300 & 3 & -23797.38 & -23815.83 & -23819.03 & -23819.15 & -23825.43 & -23824.90 \\ 
   300 & 6 & -23774.33 & -23777.53 & -23781.62 & -23777.52 & -23782.49 & -23782.12 \\ 
   500 & 1 & -25126.22 & -25388.90 & -25400.31 & -25421.05 & -25420.41 & -25420.22 \\ 
   500 & 3 & -24946.51 & -24970.60 & -24978.07 & -24975.04 & -24981.11 & -24980.70 \\ 
   500 & 6 & -24920.53 & -24925.79 & -24936.79 & -24934.07 & -24936.19 & -24934.67 \\ 
  1000 & 1 & -28835.55 & -29245.86 & -29268.02 & -29303.88 & -29287.42 & -29286.82 \\ 
  1000 & 3 & -28459.73 & -28503.98 & -28516.88 & -28512.76 & -28519.96 & -28519.22 \\ 
  1000 & 6 & -28341.42 & -28341.35 & -28353.58 & -28345.77 & -28352.02 & -28351.56 \\ 
  2000 & 1 & -37725.47 & -38356.53 & -38367.91 & -38443.95 & -38415.57 & -38415.05 \\ 
  2000 & 3 & -36795.54 & -36850.21 & -36870.57 & -36866.91 & -36874.63 & -36874.02 \\ 
  2000 & 6 & -36604.24 & -36604.09 & -36617.14 & -36613.30 & -36617.83 & -36617.49 \\ 
   \hline
\end{tabular}

\caption{Values of the DIC for the models fit to the simulated data.}
\label{table:sim}
\end{table}

\begin{figure}
\centering
\includegraphics[scale=0.5]{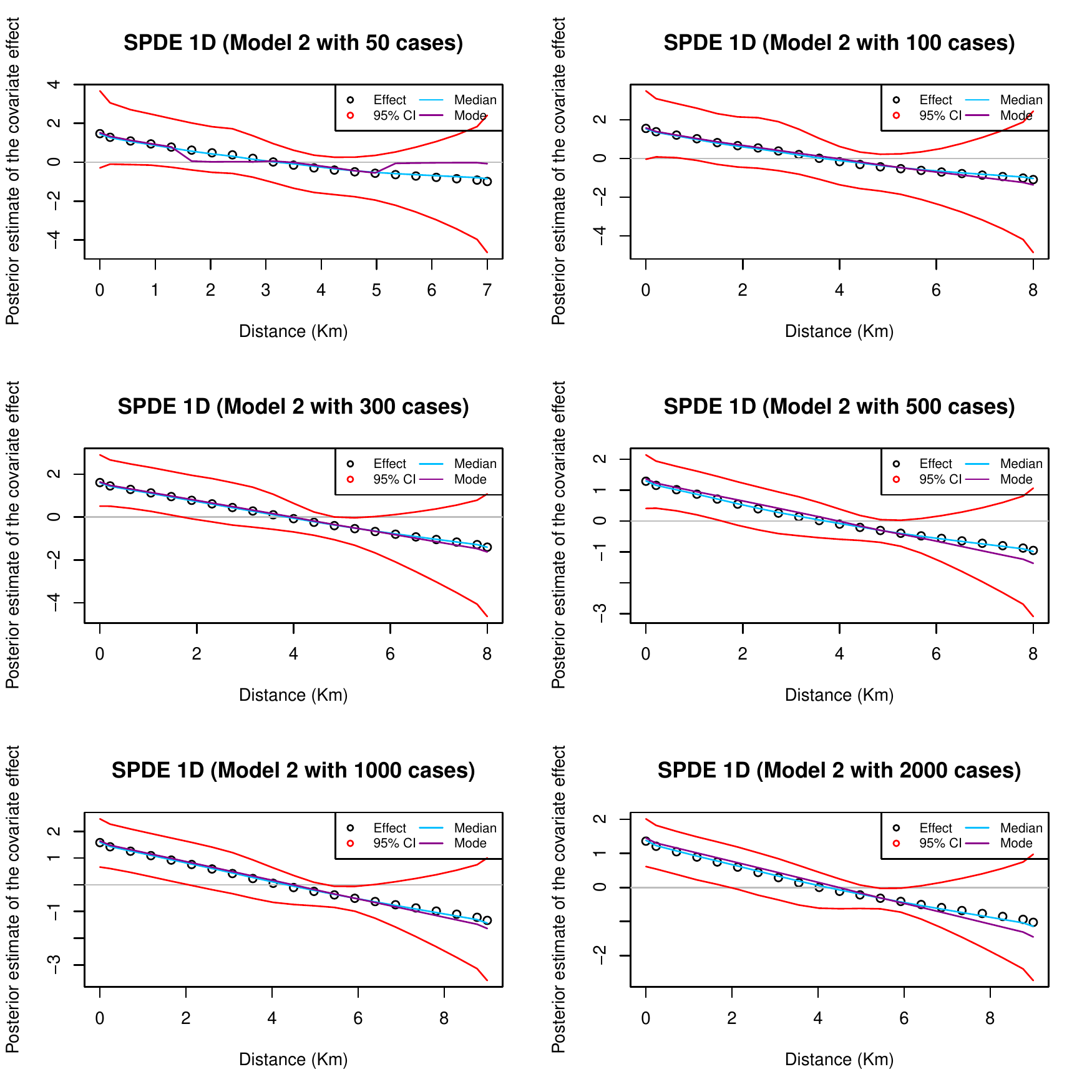}
\caption{Estimated effect of a pollution source using Model 2 and SPDE1 effects
on the simulated case-control data with $\phi$ equal to 3.}
\label{fig:sim}
\end{figure}

Figure~\ref{fig:sim} shows the estimates of the smooth terms using SPDE1
effects for different values of the sample size and $\phi$ equal to 3.  The
estimated effects are similar for RW1 effects and they are not shown. As it can
be seen, the detectability of the effects increases with the sample size, and
the credible intervals get narrower with the sample size.  Results are similar
for $\phi$ equal to 6, which produces a stronger effect and thus detecting a
significant effect requires a smaller sample size.

Model 3 did not detect exposure to the pollution source in any of the simulated
scenarios.  This is probably due to the fact that the disease-specific spatial
terms accounts for all the unexplained spatial variation.  This points to a
possible confounding between disease-specific spatial pattern $S_i(x)$ and the
effect on the covariates $\mathcal{F}_{ij}(x)$. Hence, Model 2 will be preferred
in case of doubt.

\section{Discussion}
\label{sec:discussion}

The integrated nested Laplace approximation is a suitable Bayesian inferential
framework to fit log-Gaussian Cox processes to multivariate point patterns. The log-intensity can be modeled as a sum of fixed effects and smooth terms on the
covariates plus spatial smooth terms using the SPDE approximation. Hence,
models for multivariate point patterns can be developed with ease.

These models have been applied to case-control data where cases of several
types of cancer have been considered. The models proposed in this paper have
been adequate to assess spatial risk variation and the detection of regions of
high risk. Furthermore, the assessment of risk due to the proximity to putative
pollution sources can be assessed by considering the distance from the cases to
the pollution source.  Finally, this methodology can be used to assess
differences in the disease-specific residual spatial variation between two
diseases. This can be of interest to identify diseases with a similar spatial
variation.

These models have been applied to a dataset from Alcal\'a de Henares (Madrid,
Spain) to study the spatial risk variation of lung, stomach and kidney cancer
using case-control data. These models have been able to identify a (mild)
disease-specific spatial variation for lung and stomach cancer, while the
distribution of cases of kidney cancer seems to follow that of the controls.
Models with shared and disease-specific spatial pattern have been able to
highlight the regions of high risk for lung and stomach cancer.  Furthermore,
by including the effect of the distance to important pollution sources around
the city we have been able to identify some possible sources of pollution that
affect the location of cases of lung and stomach cancer. A further
epidemiological study could look at the particular activity of each industry
and how that could be possibly linked to the increase of cancer cases.  

Although this study is limited by the small number of cases available and the
effect of the distance to the pollution source was difficult to assess, we have
carried out a simulation study that confirms that this methodology can detect
the effect of risk factors, such as exposure to pollution sources. 

In the future, we expect to extend the current methodology to perform automatic
detection of socio-economic risk factors as well as the effect of pollution
sources.  This can be done by using Reversible Jump MCMC\cite{Green:1995}
methods so that the effects on the pollution sources can be automatically
included (or removed) from the model.

This methodology can also be extended to the spatio-temporal case by modeling
the log-intensity as the sum of a spatial effect plus a temporal effect, using
a RW1 or SPDE1 effects\citep{Krainskietal:2019}. Covariates could also be considered in this model as
well.

\section{Acknowledgments}

%Anonymized data has been provided by Jos\'e Miguel Sanz Anquela, MD, PhD.
%Hospital Universitario Pr\'incipe de Asturias (Alcal\'a de Henares, Spain).

This work has been supported by grants PPIC-2014-001-P and
SBPLY/17/180501/000491, funded by Consejer\'ia de Educaci\'on, Cultura y
Deportes (JCCM, Spain) and FEDER, and grant MTM2016-77501-P, funded by
Ministerio de Econom\'ia y Competitividad (Spain).

F. Palm\'i-Perales has been supported by a Ph.D. scholarship awarded by the
University of Castilla-La Mancha (Spain).

The authors thank Mario Gonz\'alez-S\'anchez and Javier Gonz\'alez-Palacios
("Bioinformatics and Data Management Group" (BIODAMA, ISCIII)) for their
technical support in data base maintenance.

Disclaimer: This article presents independent research. The views expressed are
those of the authors and not necessarily those of the Carlos III Institute of
Health.

\bibliographystyle{SageV} 
\bibliography{Multppp}

\end{document}